\documentclass[aps,prd,twocolumn,epsf,superscriptaddress]{revtex4}

\usepackage{graphicx}
\usepackage{amsmath,amssymb,latexsym}
\usepackage{bm}

\def\lbldef#1#2{\expandafter\gdef\csname #1\endcsname {#2}}

\def\href#1#2{#2}

\newcommand{\ber}{\begin{eqnarray}}
\newcommand{\eer}{\end{eqnarray}}

\newcommand{\beqar}{\begin{eqnarray}}

\newcommand{\eeqar}{\end{eqnarray}}

\newcommand{\dsl}
  {\kern.06em\hbox{\raise.15ex\hbox{$/$}\kern-.56em\hbox{$\partial$}}}

\newcommand{\eeqarr}{\end{eqnarray}}
\newcommand{\ZZ}{{\rm \kern 0.275em Z \kern -0.92em Z}\;}

\begin{document}

\preprint{
\hfil
\begin{minipage}[t]{3in}
\begin{flushright}
\vspace*{.4in}
DESY 05-042\\
NUB--3253--Th--05\\
hep-ph/0503229
\end{flushright}
\end{minipage}
}

\title{Neutrinos as a diagnostic of cosmic ray Galactic/extra-galactic transition}

\author{Markus Ahlers}
\affiliation{Deutsches Elektronen-Synchrotron DESY,
Hamburg, Germany
}

\author{Luis A.~Anchordoqui}
\affiliation{Department of Physics,
Northeastern University, Boston, MA 02115
}

\author{Haim Goldberg}
\affiliation{Department of Physics,
Northeastern University, Boston, MA 02115
}

\author{Francis Halzen}
\affiliation{Department of Physics,
University of Wisconsin, Madison WI 53706
}

\author{Andreas Ringwald}
\affiliation{Deutsches Elektronen-Synchrotron DESY,
Hamburg, Germany
}

\author{Thomas J. Weiler}
\affiliation{Department of Physics and Astronomy,
Vanderbilt University, Nashville TN 37235
}
\date{March 2005}
\begin{abstract}

Motivated by a recent change in viewing the onset of the
extra-galactic component in the cosmic ray spectrum, we have
fitted the observed data down to $10^{8.6}$~GeV and have obtained
the corresponding  power emissivity. This transition energy is
well below the threshold for resonant $p\gamma$ absorption on the
cosmic microwave background, and thus source evolution is an
essential ingredient in the fitting procedure. Two-parameter fits
in the spectral and redshift evolution indices show that a
standard Fermi $E_i^{-2}$ source spectrum is excluded at larger
than 95\% confidence level (CL). Armed with the primordial
emissivity, we follow Waxman and Bahcall to derive the associated
neutrino flux on the basis of optically thin sources. For $pp$
interactions as the generating mechanism, the neutrino flux
exceeds the AMANDA-B10 90\%~CL upper limits. In the case of
$p\gamma$ dominance, the  flux is  consistent with AMANDA-B10
data. In the new scenario the  source neutrino flux dominates over
the cosmogenic flux at energies below $10^9$ GeV. Should data 
from AMANDA-II prove consistent with the model, we show that
IceCube can measure the characteristic power law of the neutrino
spectrum, and thus provide a window on the source dynamics.

\end{abstract}


\maketitle

\section{Introduction}

A plethora of explanations have been proposed to address the
production mechanism of ultra-high energy cosmic
rays~\cite{Anchordoqui:2002hs}. In the absence of a single model
which is consistent with all data, the origin of these particles
remains a mystery. Clues to solve the mystery are not immediately
forthcoming from the data, particularly since various experiments
report mutually inconsistent results. In recent years, a somewhat
confused picture
regarding
the energy spectrum and arrival direction distribution has been emerging.
Since 1998, the AGASA Collaboration has consistently
reported~\cite{Takeda:1998ps} a continuation of the spectrum beyond the
expected Greisen--Zatsepin--Kuzmin (GZK) cutoff~\cite{Greisen:1966jv},
which should arise at about $10^{10.7}$~GeV if cosmic ray sources are
at cosmological distances. In contrast,
the most recent results from HiRes~\cite{Abu-Zayyad:2002sf}
describe a spectrum which is consistent with the
expected GZK feature. This situation
exposes
the challenge posed by systematic errors in these types of measurements. Further
confusing the issue, the AGASA Collaboration reports observations
of event clusters which have a chance probability smaller than 1\% to arise
from a random distribution~\cite{Hayashida:bc}, whereas the recent analysis reported
by the HiRes Collaboration showed that their data are consistent with no
clustering among the highest energy events~\cite{Abbasi:2004ib}.

Deepening the mystery, recent HiRes data have been interpreted as
a change in cosmic ray composition, from heavy
nuclei to
protons,
at $\sim 10^9$~GeV~\cite{Bergman:2004bk}. This is an order of
magnitude lower in energy than the previous crossover deduced from
the Fly's Eye data~\cite{Bird:1993yi}. The end-point of the
galactic flux is expected to be dominated by iron, as the large
charge $Ze$ of heavy nuclei reduces their Larmor radius
(containment scales linearly with $Z$) and facilitates their
acceleration to highest energy (again scaling linearly with $Z$).
The dominance of nuclei in the high energy region of the Galactic
flux carries the implication that any changeover to protons
represents the onset of dominance by an extra-galactic component.
The  inference from this new HiRes data is therefore that the
extra-galactic flux is beginning to dominate the Galactic flux
already at $\sim 10^9$~GeV. Significantly, this is well below
$E_{\rm GZK} \sim 10^{10.7}$~GeV~\cite{Greisen:1966jv}, the
threshold energy for resonant $p \gamma_{\rm CMB} \rightarrow
\Delta^+ \rightarrow N \pi$ energy-loss on the cosmic microwave
background (CMB), and so samples sources even at large redshift.

The dominance of extra-galactic protons at lower energy can be
consistent with recently corroborated structures in the cosmic ray
spectrum.  A second knee, recognized originally in AGASA
data~\cite{Nagano:1991jz}, is now confirmed by the HiRes-MIA
Collaboration~\cite{Abu-Zayyad:2000ay}. At $10^{8.6}$~GeV, the
energy spectrum steepens from $E^{-3}$ to $E^{-3.3}$. This
steepening at the second knee can be
explained~\cite{Berezinsky:2002nc} by energy losses of
extra-galactic protons over cosmic distances, due to $e^+ e^-$
pair-production on the CMB. The theoretical threshold of the
energy-loss feature occurs at $10^{8.6}$~GeV, and therefore allows
for proton dominance even below this energy. However, the HiRes
data~\cite{Bergman:2004bk} seem to indicate a composition change
coincident with the energy of the second knee (from about 50\%
protons just below to 80\% protons just above), and therefore
argues for the beginning of extra-galactic proton dominance at the
second knee. Another feature in the cosmic ray spectrum is the
ankle at $\sim 10^{10}$~GeV where the spectrum flattens from
$E^{-3.3}$ to $E^{-2.7}$. This has been commonly identified with
the onset of the extra-galactic flux in the past. In the aftermath
of the new HiRes data, the ankle can now be interpreted as the
minimum in the $e^+ e^-$ energy-loss feature.

These changes in viewing the onset of the extra-galactic component
have spurred a refitting of the cosmic ray data down to
$10^{8.6}$~GeV with appropriate propagation functions and
extra-galactic injection
spectra~\cite{Berezinsky:2002nc,Abbasi:2002ta}. The major result
is that the injection spectrum is significantly steeper than the
standard $E_i^{-2}$ predicted by Fermi engines. This result has
consequences for neutrino observation: predictions for both the
cosmogenic fluxes (produced via interactions of super-GZK
cosmic-rays on the CMB) and the direct neutrino luminosity from
optically thin sources can be significantly modified. The
implication for cosmogenic neutrinos has been discussed
elsewhere~\cite{Fodor:2003ph,Seckel:2005cm}. In this paper we
analyze the impact on neutrino luminosities from optically thin
sources which are associated with this change in view of the
Galactic/extra-galactic crossover energy.

The outline of the paper is as follows. We begin in
Sec.~\ref{EGCRP} with an estimate of the power density of cosmic
rays assuming a ``low'' energy  onset of dominance by an
extra-galactic component. This is accomplished by
a goodness-of-fit test of our scenario with
the energy spectrum as observed by the Akeno~\cite{Nagano:1991jz} +
AGASA~\cite{Takeda:1998ps} and the Fly's
Eye~\cite{Bird:1993yi,Bird:1994wp} +
HiRes~\cite{Abu-Zayyad:2002sf} experiments, respectively, in the
energy range from the second knee upward, {\em i.e.,}
$10^{8.6}$~GeV $< E <$ $10^{11}$~GeV. In the fitting procedure we
use appropriate propagation functions~\cite{Fodor:2000yi} that
take into account photo-meson and pair production on the CMB.
Armed with the cosmic ray emissivities required to populate the
observed spectrum with extra-galactic protons all the way down to
$10^{8.6}$~GeV, in Sec.~\ref{NCR} we derive predictions of
neutrino fluxes associated with $p\gamma$ or $pp$ interactions in
sources which are optically thin. To this end, we follow the
procedure delineated by Waxman and Bahcall
(WB)~\cite{Waxman:1998yy}, but instead of assuming a specific
cosmic ray injection spectrum $\propto E_i^{-2}$ and redshift
source evolution $\propto (1 + z)^3,$ we use values for the
spectral and source  evolution indices complying with the best
fits to the spectra obtained in Sec.~\ref{EGCRP}. In Sec.~\ref{CN}
we  review the computation of cosmogenic neutrinos and show that
in the ``new'' low crossover scenario, associated neutrino spectra
from optically thin sources  dominates over the cosmogenic flux.
In Sec.~\ref{icecube} we calculate event rates at
IceCube~\cite{Ahrens:2003ix} expected from source fluxes derived
in Sec.~\ref{NCR}, and on this basis assess the potential of this
detector to constrain the crossover energy between Galactic and
extra-galactic dominance in the cosmic ray spectrum.
Section~\ref{conclusions} contains a summary of our results and
conclusions.

\section{Extra-galactic cosmic ray power}
\label{EGCRP}

It is helpful to envision the cosmic ray engines as machines where
protons are accelerated and (possibly) permanently confined by the
magnetic fields of the acceleration region. The production of
neutrons and charged pions and subsequent decay produces both
neutrinos and cosmic rays: the former via $\pi^+ \rightarrow e^+
\nu_e \nu_\mu \overline \nu_\mu$ (and the conjugate process), the
latter via neutron diffusion from the region of the confined
protons. If the neutrino-emitting source  also produces high
energy cosmic rays, then  pion production must be the principal
agent for the high energy cutoff on the proton spectrum.
Conversely, since the protons must undergo sufficient
acceleration, inelastic pion production needs to be small below
the cutoff energy; consequently,  the plasma must be optically
thin. Since the interaction time for protons is greatly increased
over that of neutrons because of magnetic confinement, the
neutrons escape before interacting, and on decay give rise to the
observed cosmic ray flux. The foregoing can be summarized as three
conditions on the characteristic nucleon interaction time scale
$\tau_{\rm int}$; the neutron decay lifetime $\tau_n$;  the
characteristic cycle time of confinement $\tau_{\rm cycle}$; and
the total proton confinement time $\tau_{\rm conf}$: $(1)\;
\tau_{\rm int}\gg \tau_{\rm cycle}$; $(2)\; \tau_n > \tau_{\rm
cycle}$; $(3)\; \tau_{\rm int}\ll \tau_{\rm conf}.$ The first
condition ensures that the protons attain sufficient energy.
Conditions $(1)$ and $(2)$ allow the neutrons to escape the source
before decaying. Condition $(3)$ permits sufficient interaction to
produce neutrons and neutrinos. We take these three conditions
together to define an optically thin source. A desirable property
of this low-damping scenario is that a single source will produce
cosmic rays with a smooth spectrum across a wide range of energy.

\begin{figure}
\begin{center}
\includegraphics[height=5.55cm]{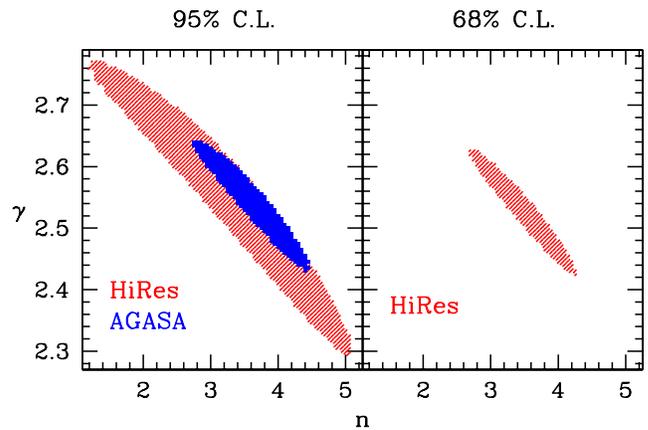}
\caption{
The goodness-of-fit test of the low crossover scenario, using the
method outlined in Ref.~\cite{Fodor:2003ph}.
In the left panel we show the 95\% CL allowed regions in
the $n-\gamma$ plane, obtained with Fly's Eye + HiRes and  Akeno +
AGASA data. In the right panel we show the corresponding 68\% CL
obtained from Fly's Eye + HiRes data.  In  the fitting procedure
we used data in the energy interval
from $E_- = 10^{8.6}$~GeV to $E_+ = 10^{11}$~GeV,
taking $z_{\rm min} = 0.012$ and $z_{\rm max} =
2.$ } \label{gxg_cl}
\end{center}
\end{figure}

Assigning extra-galactic dominance to energies beginning at $\sim
10^{9}$~GeV, rather than $\sim 10^{10}$~GeV, increases the
required energy production rate of extra-galactic cosmic rays. The
power density in the energy range $10^{10}$~GeV to $10^{12}$~GeV
is found to be $\dot \epsilon_{_{\rm CR}} [10^{10}, 10^{12}] \sim
5 \times 10^{44}$~erg Mpc$^{-3}$ yr$^{-1}$~\cite{Waxman:1995dg}.
As emphasized in~\cite{Waxman:1995dg}, this result is independent
of source evolution: for the stated energy interval, cosmic rays
from distant sources will undergo significant energy losses on the
CMB, and thus only nearby sources contribute to the observed
spectrum. In what follows we obtain analogous power densities
corresponding to the lower energy onset of extra-galactic
dominance. As can be expected, these will have sensitivity to
source evolution.

We assume an isotropic distribution of neutron-emitting sources
that can be described by a comoving luminosity distribution
${\cal L}_n (r, E_i)$ where $E_i$ is the injection
energy and $r$ the distance to Earth.
The number of protons $N_p$ arriving at Earth with energy $E$ per units of energy, area $A$,
time $t$ and solid angle $\Omega$ reads,
\begin{eqnarray}
J_p & \equiv & \frac{d^4N_p}{dE\,dA\,dt\,d\Omega} \nonumber \\
     & = & \frac{1}{4\pi} \int_0^\infty dE_i \int_0^\infty dr
\left| \frac{\partial P_{p|n} (E;E_i, r)}{\partial E}
\right|\,{\cal L}_n \,\,, \label{propagation}
\end{eqnarray}
where $P_{p|n}$ is the propagation function introduced in Ref.~\cite{Fodor:2000yi} which gives the expected
number of protons above a threshold energy $E$ if a neutron with energy $E_i$
was emitted from a source at a distance $r$. The Monte-Carlo program that computes $P_{p|n}$
uses SOPHIA~\cite{Mucke:1999yb} as an external package for simulation of the GZK interactions.
The  program  exploits
 a continuous energy loss approximation to describe the $e^+e^-$ pair production process.
To estimate the differential flux of protons, we calculate the $P_{p|n}$ function
for infinitesimal steps ($1 \div 10$ kpc) as a function of the redshift $z$ and multiply the
corresponding infinitesimal probabilities starting at a distance $r(z)$ down to Earth with $z=0.$

We  take all sources to have identical
injection spectra $d\dot N_n/dE_i\propto E_i^{-\gamma}\,
\Theta (E_{i, {\rm max}}- E_i)$, and parametrize the redshift
evolution of the source luminosity and  the  comoving
number density  $\rho_{_{\rm CR}}$ by a simple power-law,
\begin{equation}
\label{emissivity_distr}
   {\cal L}_n = \rho_{_{\rm CR}}\, \left[ 1 + z (r) \right]^n\,
   \Theta (z - z_{\rm min})\, \Theta (z_{\rm max} - z)\, \frac{d\dot N_n}{dE_i}\, ,
\end{equation}
where the redshift $z$ and the distance $r$ are related by $d z =
(1 + z)\, H (z)\, d r$. The Hubble expansion rate at a redshift
$z$ is related to the present one $H_0$ through $H^2 (z) = H^2_0\,
\left[\Omega_M (1 + z)^3 + \Omega_{\Lambda} \right],$ where
$\Omega_M$ and $\Omega_\Lambda$ are the matter and vacuum-energy
densities in terms of the critical density. Here we take $\Omega_{M} = 0.3$ and $\Omega_{\Lambda} =
0.7,$ in agreement with WMAP observations~\cite{Spergel:2003cb}.
The results turn out to be rather insensitive to the precise
values of the cosmological parameters within their uncertainties.
The minimal and maximal redshift parameters $z_{\rm min}$ and
$z_{\rm max}$ exclude the existence of nearby and early time
sources. As default value, we take $z_{\rm min}=0.012$, corresponding
to $r_{\rm min}=50$~Mpc, and comment on the effect of possible
variations where appropriate.
Note that the effects due to a
change in $z_{\rm max}$ can be largely compensated by a change in
$n.$
Therefore, we fix $z_{\rm max}=2$ in the following and study the
dependences on $n$ only.
For the maximum injection energy we take as a default value
$E_{i,{\rm max}}=10^{12.5}$~GeV. Most of our results are insensitive to
this choice, as long as $E_{i,{\rm max}}$ is above $\sim 10^{11.5}$~GeV
(see also Ref.~\cite{Fodor:2003ph}), as we will see in the following.

\begin{figure}
\begin{center}
\includegraphics[height=8.4cm]{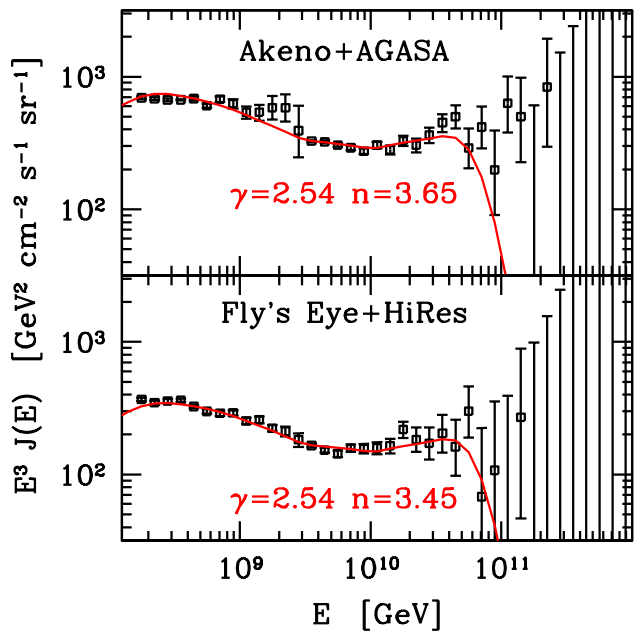}
\caption{Best fits to the ultra-high energy cosmic ray spectrum
in the energy interval $[E_-,\; E_+]$ as observed by Akeno + AGASA and
Fly's Eye + HiRes. We set
$E_- = 10^{8.6}$~GeV,
$E_+ = 10^{11}$~GeV,
$z_{\rm min} = 0.012,$
$z_{\rm max} = 2,$
and
$E_{i,{\rm max}} = 10^{12.5}$~GeV.}
\label{gxg_fitspectra}
\end{center}
\end{figure}

The  cosmic ray spectra as observed by
Akeno~\cite{Nagano:1991jz} +
AGASA~\cite{Takeda:1998ps}
and
Fly's Eye~\cite{Bird:1993yi,Bird:1994wp} +
HiRes~\cite{Abu-Zayyad:2002sf} are fitted and confidence
levels are assigned using a Poisson likelihood following the
procedure detailed in Ref.~\cite{Fodor:2003ph}. The 95\% CL
exclusion contours in the $n - \gamma$ plane for the two data
samples are shown in Fig.~\ref{gxg_cl}. The parameters for the
best fit, shown in Fig.~\ref{gxg_fitspectra}, are given in
Table~\ref{t1}. The disparity in the
goodness-of-fit tests of the two
data samples is largely originating in the presence of a spurious
bump in the region of $10^{9.4}$ GeV of the Akeno + AGASA data
(cf. Ref.~\cite{Fodor:2003ph}).
This in turn stems from combining the data of the Akeno array and
the full AGASA experiment.  Interestingly,
{\it the new scenario
with extra-galactic cosmic rays dominating the spectrum below the
ankle, down to the second knee at $E_- = 10^{8.6}$~{\rm GeV}, is inconsistent
at more than a $2 \sigma$ level
with standard Fermi engine models that suggest an injection
spectrum $\propto E_i^{-2}$.} Note that for both data samples the
best fit yields $\gamma = 2.54$. Additionally, $\gamma < 2.4$ is
disfavoured at the $1\sigma$ level by
Fly's Eye~\cite{Bird:1993yi,Bird:1994wp} + HiRes~\cite{Abu-Zayyad:2002sf}
data  and
at the $2\sigma$ level by
Akeno~\cite{Nagano:1991jz} + AGASA~\cite{Takeda:1998ps}
data. We have checked that
this is robust against variations of $z_{\rm min}\leq 0.012$ and
$E_{i,{\rm max}}\geq 10^{11.3}$~GeV.

It is conceivable that the extra-galactic proton flux
largely exceeds the
nearby data below $10^{8.6}$ GeV. As can be seen in
Fig.~\ref{gxg_fitspectra}, this is not the case for the best-fit
values. {\em Moreover, if we make a more sophisticated analysis with a
Galactic component below $10^{8.6}$ GeV, the 2$\sigma$ allowed
regions will shrink.}

\begin{table}
\caption{Best fit parameters}
\begin{center}
\begin{tabular}{c|c|c|c}
\hline \hline
Experiment & $\gamma$  & $n$  &
$\dot \epsilon_{_{\rm CR}} [10^{8.6}\ \mathrm{GeV}, 10^{12.5}\ \mathrm{GeV}]
$ \\ \hline
AGASA & $\,\,\,\, 2.54 \,\,\,\,$ & $\,\,\,\, 3.65 \,\,\,\,$ &
$\, 2.5 \times 10^{45}$~erg Mpc$^{-3}$ yr$^{-1}$ \\
HiRes & 2.54 & 3.45 & $\, 1.3 \times 10^{45}$~erg Mpc$^{-3}$ yr$^{-1}$ \\
\hline \hline
\end{tabular}
\end{center}
\label{t1}
\end{table}

At this stage, it is worthwhile to mention that we have verified the consistency
of the simulations by fitting the data above $10^{10}$~GeV and comparing the
cosmic ray power density,
\begin{equation}
\dot \epsilon_{_{\rm CR}} [E_{i, {\rm  min}},\, E_{i, {\rm max}}]=\rho_{_{\rm CR}}\,
\int_{E_{i,\mathrm{min}}}^{E_{i,\mathrm{max}}}
\,dE_i\,E_i\,\,\frac{d\dot N}{dE_i}.
\end{equation}
with the result obtained in~\cite{Waxman:1995dg}.
Our best fits, using
Eq.~(\ref{propagation}) with $\gamma =2$ and $n=3$, correspond to
$\dot \epsilon_{_{\rm CR}} [10^{10}\ \mathrm{GeV}, 10^{12}\ \mathrm{GeV}]
= 4.8 \times 10^{44}$~erg Mpc$^{-3}$ yr$^{-1}$
and $\dot \epsilon_{_{\rm CR}} [10^{10}\ \mathrm{GeV}, 10^{12}\ \mathrm{GeV}]
= 2.4 \times 10^{44}$~erg Mpc$^{-3}$ yr$^{-1}$,
for AGASA and HiRes, respectively.

\section{Neutrino $\bm{\rightleftharpoons}$ cosmic ray connection}
\label{NCR}

For optically thin sources, the neutrino power density scales
linearly with the cosmic ray power density $\dot \epsilon_{_{\rm
CR}}$~\cite{Waxman:1998yy}. The actual value of the neutrino flux
depends on what fraction of the proton energy is converted to
charged pions (which then decay to neutrinos). To quantify this,
we follow WB and define $\epsilon_\pi$ as the ratio of charged
pion energy to the {\em emerging} nucleon energy at the source.
Depending on the relative ambient gas and photon densities,
charged pion production proceeds either through inelastic $pp$
scattering~\cite{Anchordoqui:2004eu}, or photopion production
predominantly  through the resonant  process $p\gamma \rightarrow
\Delta^+\rightarrow n\pi^+$~\cite{Waxman:1998yy}. For the first of
these, the inelasticity of the process is
0.6~\cite{Alvarez-Muniz:2002ne}. This then implies that the energy
carried away by charged pions is about equal to the emerging
nucleon energy, yielding (with our definition)
$\epsilon_\pi\approx 1.$ For resonant photoproduction, the
inelasticity is kinematically determined by requiring equal boosts
for the decay products of the $\Delta^+$~\cite{Stecker:1968uc},
giving $\epsilon_\pi = E_{\pi^+}/E_n \approx 0.28$, where
$E_{\pi^+},\;E_n$ are the emerging charged pion and neutron
energies, respectively.

In this section, we will extend the WB
analysis~\cite{Waxman:1998yy} to the case of a lower onset of the
extra-galactic component. The present analysis differs from WB in
that the integrated power spectrum has changed, and that the
spectral index $\gamma \ne 2.$ We will restrict the ensuing
discussion to the case of photopion production on resonance, and
comment on the $pp$ possibility at the end of the section.

The intermediate state of the reaction $p + \gamma \to N + \pi$ is
dominated by the $\Delta^+$ resonance. In order to normalize to
the observed cosmic rays, we restrict our interest to the $n\pi^+$
decay channel. Each $\pi^+$ decays to 3 neutrinos and a positron,
$\pi^+ \to \mu^+ \nu_\mu \to \nu_\mu \overline \nu_\mu \nu_e e^+$.
The $e^+$ readily loses its energy through synchrotron radiation
in the source magnetic fields. The average  neutrino energy from
the direct pion decay is $\langle E_{\nu_\mu} \rangle_\pi =
(1-r)\,E_\pi/2 \simeq 0.22\,E_\pi$ and that of the muon is
$\langle E_{\mu} \rangle_\pi = (1+r)\,E_\pi/2 \simeq 0.78\,E_\pi$,
where $r$ is the ratio of muon to the pion mass squared. Now,
taking the $\nu_\mu$ from muon decay to have 1/3 the energy of the
muon, the average energy of the $\nu_\mu$ from muon decay is
$\langle E_{\nu_\mu} \rangle_\mu =(1+r)E_\pi/6=0.26 \, E_\pi$.
This means that neutrinos carry away about 3/4 of the $\pi^+$
energy, and each neutrino carries a fraction $\epsilon_\pi/4$ of
the accompanying cosmic ray energy.

\begin{figure}
\begin{center}
\includegraphics[height=6.6cm]{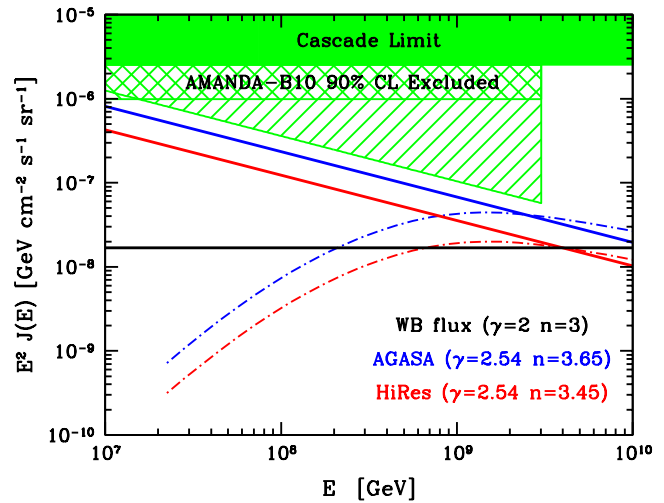}
\caption{Neutrino fluxes (summed over all flavors) from optically
thin sources for $\epsilon_\pi = 0.28.$ The horizontal solid line
indicates the WB prediction which corresponds to a
Galactic/extra-galactic crossover energy at the ankle, $\sim
10^{10}$ GeV. The falling solid lines indicate the expected
neutrino flux normalized to HiRes (lower) and AGASA (upper) data,
if one assumes the onset of dominance by the extra-galactic
component is at $10^{8.6}$~GeV. The dash-dotted lines indicate the
fluxes of cosmogenic neutrinos associated with flux predictions
given by the falling solid lines. The cross-hatched region
excludes an $E^{-2}$ spectrum at the 90\% CL by measurements of
AMANDA-B10~\cite{Ackermann:2005sb}.  The single hatched region,
obtained by rescaling the AMANDA integrated bolometric flux limit
to an $E^{-2.54}$ power law, is the exclusion region for the low
crossover model. The shaded region indicates the cascade limit
(see text for details).} \label{gxg_flux}
\end{center}
\end{figure}

In order to correlate the neutrino and  cosmic ray fluxes, we
assume both follow a common power law at injection
\begin{equation}
\frac{d \dot N_i}{dE_i} = C_{{\rm CR}(\nu)} \, E_i^{-\gamma} \,\,
\end{equation}
and normalize our spectrum in a bolometric fashion
$$
C_\nu \int_{\epsilon_\pi E_1/4}^{\epsilon_\pi E_2/4} E_i^{-(\gamma
- 1)} dE_i = \frac{3}{4}\epsilon_\pi C_{\rm CR} \int_{E_1}^{E_2}
 E_i^{-(\gamma - 1)} dE_i\,\, .
$$
After integration we have,
\begin{equation}
C_\nu \left(\frac{\epsilon_\pi}{4}\right)^{-(\gamma -2)} =
\frac{3}{4}\epsilon_\pi\; C_{\rm CR}\,\,.
\end{equation}
Therefore, for the ``low'' crossover energy scenario, the
resulting flux of neutrinos (all flavors) from optically thin
sources is given by~\cite{Waxman:1998yy}
\begin{widetext}
\begin{equation}
J_\nu (E) = 3\,\left(\frac{\epsilon_\pi}{4}\right)^{\gamma-1}
 \frac{\rho_\mathrm{CR}}{4 \pi} \left.\frac{d\dot
N_n}{dE_i}\right|_{E_i=E}\,\, \int_{z_{\rm min}}^{z_{\rm max}}
\frac{ (1+z)^{(n-\gamma)}}{H(z)}  dz  \sim 3.5 \times 10^{-3}
\left(\frac{E}{{\rm GeV}}\right)^{-2.54}~{\rm GeV}^{-1} {\rm
cm}^{-2} {\rm s}^{-1}  {\rm sr}^{-1} , \label{newWB}
\end{equation}
\end{widetext}
where the numerical value is an average flux obtained from best
fits to AGASA and HiRes data, derived in
Sec.~\ref{EGCRP}~\cite{Eberle:2004ua}. This extends the WB analysis to values
of $\gamma\ne 2.$ The fluxes for each of the
best fits are shown in Fig.~\ref{gxg_flux}, along with the WB
flux. Also shown is the region excluded by AMANDA-B10 for both the
cases $\gamma=2$ and $\gamma=2.54$, and the cascade
limit~\cite{Berezinsky:1975} from Ref.~\cite{Mannheim:1998wp},
which applies to all scenarios where neutrinos originate from pion
decays~\cite{Sreekumar:1997un}. For the low-crossover scenario,
the neutrino flux associated with  the AGASA data set is
consistent with the AMANDA-B10 data. Sensitivity to this flux will
be attained by a full analysis of the AMANDA-II data
set~\cite{francis}. In the event of a positive indication by
AMANDA-II, the IceCube facility will (as shown in
Sec.~\ref{icecube}) be capable discriminating between the low and
high crossover scenarios.

Several other remarks are in order. $(a)$ The best fit source
neutrino fluxes in Fig.~\ref{gxg_flux} correspond to $z_{\rm
min}=0.012$ and $E_{i,{\rm max}}=10^{12.5}$~GeV. We have checked
that a change of $z_{\rm min}$ to zero and a variation of
$E_{i,{\rm max}}$ within the range $[10^{11.3}\ {\rm
GeV},10^{13.5}\ {\rm GeV}]$ produces changes within the order of
the thickness of the lines. $(b)$ Because of oscillations, the
neutrino flavor mix at the source will evolve to a 1:1:1 mix at
the detector~\cite{Beacom:2003nh}. $(c)$ Electron antineutrinos
can also be produced through neutron $\beta$-decay. However, this
contribution turns out to be negligible (about 3 orders of
magnitude smaller than the charged pion
contribution)~\cite{Anchordoqui:2004eb}. $(d)$ The rapid rise at
lower energies of the low crossover neutrino spectrum will greatly
increase the event rate at the Glashow resonance as compared to
that given in~\cite{Anchordoqui:2004eb} based on the WB flux. The
$\overline\nu_e$ flux at the Glashow resonance obtained in the low
crossover scenario is about an order of magnitude smaller than the
bound from AMANDA-II data~\cite{Ackermann:2004zw}. $(e)$ Neutrino
fluxes from nearby isolated sources ({\em e.g.,} Centaurus
A~\cite{Anchordoqui:2004eu,Anchordoqui:2001nt} or
Cygnus-OB2~\cite{Anchordoqui:2003vc}) can be distinguished through
their point anisotropies, from the differential flux derived in
this work.

We now comment on the $pp$ scenario. If the primary proton
spectrum $\propto E_i^{-\gamma},$ the dominance of inelastic $pp$
collisions produces an
isotropically neutral mix of pions that on
decay give rise to a neutrino flux with spectrum $\propto
E^{-\gamma}$~\cite{Anchordoqui:2004bd}. Current hadronic event
generators yield an inelasticity of $\sim
0.6$~\cite{Alvarez-Muniz:2002ne} for the reaction $pp\rightarrow
NN + {\rm pions},$ where the $N$'s are final state nucleons. With
our definition, $\epsilon_\pi\approx 1,$ assuming that 2/3 of the
final state pions are charged. Then, the correction due to a
larger inelasticity of $pp$ interactions as compared to the
resonant $p\gamma$ scattering (with $\epsilon_\pi\approx 0.28$)
would increase the neutrino flux predictions given in
Fig.~\ref{gxg_flux} by a factor of $\approx 7$.  {\em This
sizeable augmentation of the neutrino flux based on optically thin
sources with dominant $pp$ scattering will result in the
exclusion of the low crossover scenario.} Therefore, we will
continue to present our results on the basis of the dominance of
the photopion process.

\begin{figure}
\begin{center}
\includegraphics[height=8.4cm]{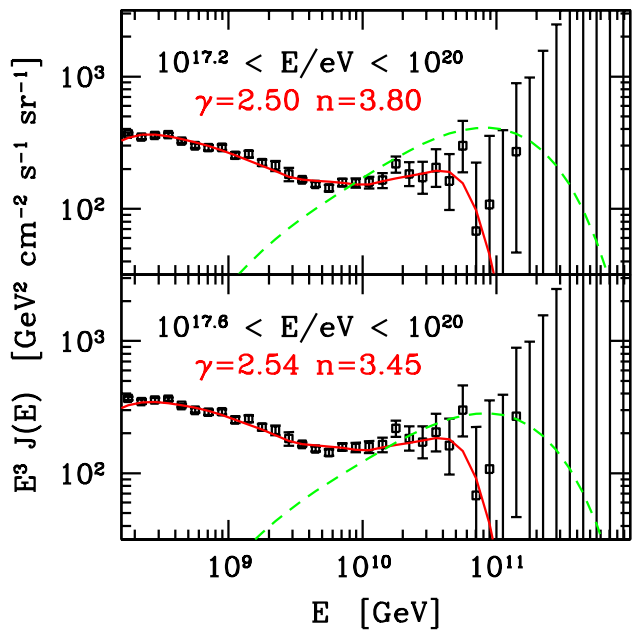}
\caption{Best fit to HiRes data (solid), assuming dominance of
the extra-galactic component above $E_- = 10^{8.2}$~GeV (top) and
$E_- = 10^{8.6}$~GeV (bottom). Also shown
is the associated cosmogenic neutrino flux for all flavors (dashed).}
\label{gxg_cosmogenic}
\end{center}
\end{figure}

\section{Cosmogenic Neutrinos}
\label{CN}

The opacity of the CMB to ultra-high energy protons propagating
over cosmological distances guarantees a cosmogenic flux of
neutrinos, originated in the reaction $p + \gamma_{\rm CMB} \to N
+ \pi$~\cite{Berezinsky:1969}. Very recently, one of us has
performed an investigation of the actual size of the cosmogenic
neutrino flux~\cite{Fodor:2003ph} assuming that all observed
cosmic ray showers above $10^{8.2}$~GeV are initiated by  protons,
with sources  isotropically distributed throughout the universe.
The low energy cutoff used in~\cite{Fodor:2003ph}  is near the
crossover energy suggested by HiRes data, and thus we expect no
significant modification on the prediction of cosmogenic neutrinos. To verify
this assertion, we estimate the flux of neutrinos produced as
sub-products in the GZK chain reaction by the population of
protons that best reproduces the HiRes data. Such a cosmogenic
flux is obtained by replacing  $P_{p|n}$ in
Eq.~(\ref{propagation}) with  $P_{\nu|n}$,
\begin{equation}
J_{\nu}
      =  \frac{1}{4\pi} \int_0^\infty dE_i \int_0^\infty dr
\left|\frac{\partial P_{\nu|n} (E;E_i, r)}{\partial E} \right|\,{\cal L}_n \,\,.
\end{equation}
In Fig.~\ref{gxg_cosmogenic} it is seen that the cosmogenic flux
predictions for the low energy crossovers at $10^{8.2}$~GeV and
$10^{8.6}$~GeV are compatible within errors. It should be noted
that the contribution to the cosmogenic neutrino flux resulting
from neutron beta decay is negligible for the energies under
consideration~\cite{Engel:2001hd}.

The cosmogenic neutrino flux corresponding to the standard
$E_i^{-2}$ injection spectrum has been previously obtained in
Ref.~\cite{Engel:2001hd} for several assumed source evolution
indices. A comparison with the WB flux shows that these (the
cosmogenic and source fluxes) are comparable at energies above
$10^8$~GeV. This is in striking contrast with the fluxes resulting
from the low crossover scenario. As can be seen in
Fig.~\ref{gxg_flux}, the source flux dominates the cosmogenic flux
at  energies below $10^9$ GeV in the low crossover
scenario~\cite{paramcosm}. Thus, the neutrinos below this energy
behave as ``unscathed messengers'' of the source injection
spectrum. The observation of a neutrino flux with a power law
spectral index $> 2.4$ can provide strong support for the low
crossover scenario.

\section{IceCube sensitivity}
\label{icecube}

In the previous sections we have shown that if the
nucleon-emitting sources are optically thin, then the diffuse flux
of neutrinos produced by these sources provides a powerful tool in
discriminating between Galactic/extra-galactic cosmic ray origin.
Should the entire scenario not be ruled out by AMANDA-II data, it
is of interest to explore the potential of forthcoming neutrino
telescopes to provide conclusive identification of the crossover
energy.

In deep ice/water/salt, neutrinos are detected by observation of
the \v {C}erenkov light emitted by charged particles produced in
neutrino interactions. In the case of an incident high-energy muon neutrino,
for instance, the neutrino interacts with a hydrogen or oxygen
nucleus in the deep ocean water (or ice) and produces a muon
travelling in nearly the same direction as the neutrino. The blue
\v {C}erenkov light emitted along the muon's kilometer-long
trajectory is detected by strings of photomultiplier tubes
deployed at depth shielded from radiation. The orientation of the
\v {C}erenkov cone reveals the neutrino direction. There may also
be a visible hadronic shower if the neutrino is of sufficient
energy.

The Antarctic Muon And Neutrino Detector Array
(AMANDA)~\cite{Andres:1999hm}, using natural 1 mile deep Antarctic ice as a \v {C}erenkov detector,
has operated for
more than 3 years in its final configuration of 680 optical modules on 19 strings. The detector is in
steady operation collecting roughly four neutrinos per day using fast on-line analysis software. Its
performance has been calibrated by reconstructing muons produced by atmospheric muon
neutrinos~\cite{Andres:2001ty}.

\begin{figure}
\begin{center}
\includegraphics[height=10.3cm]{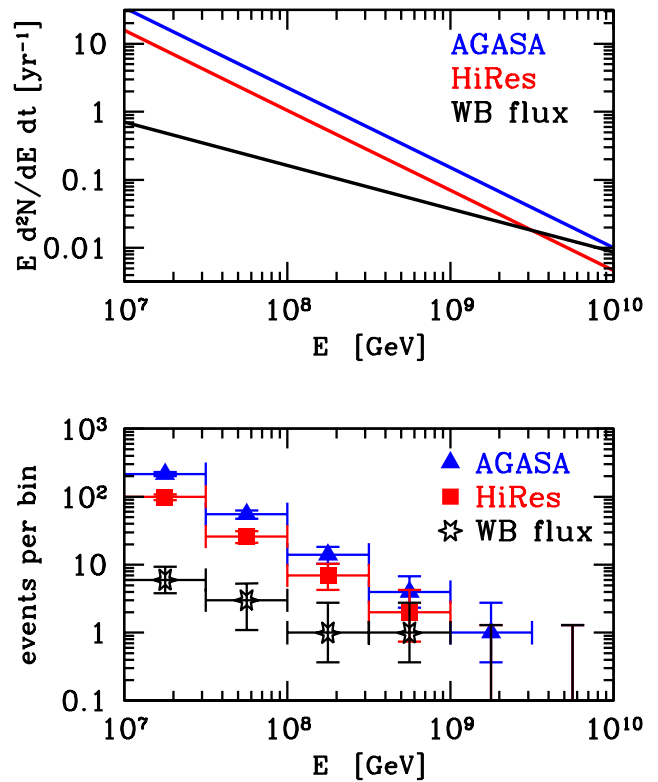}
\caption{Upper panel: Differential event rate at IceCube for the different
neutrino flux predictions from AGASA (top), HiRes (middle) and WB (bottom)
obtained in Sec.~\ref{NCR}. Lower panel: Expected bin-by-bin event rates for 10 years of operation.
The bin  partition interval is taken as $\Delta \log_{10} E = 0.5.$}
\label{gxg_icecube}
\end{center}
\end{figure}

Overall, AMANDA represents a proof of concept for the kilometer-scale neutrino observatory,
IceCube~\cite{Ahrens:2003ix}, now under construction. IceCube will consist of 80 kilometer-length strings,
each
instrumented with 60 10-inch photomultipliers spaced by 17~m.
The deepest module is 2.4~km below the surface. The strings are
arranged at the apexes of equilateral triangles 125\,m on a side. The instrumented (not effective!)
detector volume is a cubic kilometer. A surface air shower detector, IceTop, consisting of 160
Auger-style~\cite{Abraham:2004dt} \v {C}erenkov detectors deployed over 1\,km$^{2}$ above IceCube,
augments the deep-ice
component by providing a tool for calibration, background rejection and air-shower physics.
Construction of the detector started in the Austral summer
of 2004/2005 and will continue for 6 years, possibly less. At the time of writing, data
collection by the first string has begun.

At the energies under consideration, there is no
atmospheric muon or neutrino background in a
km$^3$ detector.
The differential event rate  is given by
\begin{equation}
\frac{d^2N}{dE\,dt} \approx 2 \pi N_A \,\,\, \rho \,\, V_{\rm eff} \,
J_{\nu}\,\,\,
\sigma^{{\rm CC}}_{\nu N} \,\,\,  ,
\label{cenaice}
\end{equation}
where $N_A$ is Avogadro's number, $V_{\rm eff} \approx 2$~km$^3$
is the effective volume of ice with density $\rho,$ and
$\sigma^{{\rm CC}}_{\nu N}=6.78\times 10^{-35}\ (E/{\rm
TeV})^{0.363}$~cm$^2$ is the charged current  neutrino-nucleon
cross section~\cite{Gandhi:1998ri}. The effective volume used is
conservative, since muon tracks can originate well outside the
fiducial volume of the detector~\cite{Halzen:2002pg}. In
Fig.~\ref{gxg_icecube} we show the differential event rate at
IceCube from optically thin sources. Also shown are the expected
bin-by-bin event rates for 10 years of data collection, with a bin
partition size $\Delta \log_{10} E = 0.5.$ The vertical error bars
are obtained on the basis of Poisson statistics with $\Delta
\log_{10} N= 0.434 \sqrt{N}/N,$ for $N>20$. For smaller statistics
we use Poisson confidence intervals~\cite{Feldman:1997qc}. It is
strongly indicated that within its lifetime IceCube will attain
sufficient sensitivity to constrain the energy of transition
between Galactic and extra-galactic dominance.
RICE~\cite{Kravchenko:2001id}, PAO~\cite{Capelle:1998zz},
EUSO~\cite{Scarsi:2001fy}, ANITA~\cite{Silvestri:2004uw}, and
OWL~\cite{Stecker:2004wt} also have the potential to measure the
ultra-high energy neutrino flux. However, the energy thresholds,
systematics, backgrounds, or time-scales to completion leave
 these experiments less promising than IceCube for a spectrum
determination in the near future.

\section{Conclusions}
\label{conclusions}

We have estimated the extragalactic diffuse neutrino flux emitted from optically
thin sources, on the basis of a low
transition energy $(10^{8.6}$ GeV) between Galactic and extragalactic cosmic rays.
Such a low crossover finds
support in the chemical composition analysis of HiRes data~\cite{Bergman:2004bk}, and is sustained by
studies which reproduce the steepening at the second knee via $e^+e^-$ production on the
CMB~\cite{Berezinsky:2002nc}. Since the neutrino flux reflects the nucleon flux at the source, the latter
must be obtained by fitting observed cosmic ray data taking into account propagation
effects. The low crossover energy is well below the threshold energy for resonant
$p\gamma_{\rm CMB}$ absorption, and so samples sources even at large redshift. Thus,
source evolution is an important consideration in this calculation. Two-parameter
fits in the spectral and redshift evolution indices show that a standard Fermi $E_i^{-2}$
source spectrum is excluded at larger than 95\% CL. Best fits to both Akeno~\cite{Nagano:1991jz} +
AGASA~\cite{Takeda:1998ps} and
Fly's Eye~\cite{Bird:1993yi,Bird:1994wp} + HiRes~\cite{Abu-Zayyad:2002sf}
data sets give an $E_i^{-2.54}$ source spectrum, with an evolution index somewhat larger
than 3. The neutrino flux  obtained using the WB~\cite{Waxman:1998yy} consideration for energetics
at the source mirrors the steep spectrum of the emitted cosmic rays.

Comparison of the resulting flux with existing AMANDA-B10 90\%~CL
bounds~\cite{Ackermann:2005sb} reveals the following: (1) If
neutrinos are generated by $pp$ interactions at the source, the
resulting flux is within the excluded region. (2) For $p\gamma$
interactions dominant, the best fit to the data yields a neutrino
flux which is consistent with the AMANDA-B10 upper limit. A
complete analysis of the AMANDA data will provide sufficient
sensitivity to rule out the model.

The neutrino flux at the source in this scenario dominates the cosmogenic flux. Thus,
should data from AMANDA-II not rule out the model, we show that IceCube can measure
the characteristic power law of the neutrino spectrum, and thus provide a window on the
source dynamics.

In summary, forthcoming data from the South Pole can provide significant clues in
demarcating the cosmic ray Galactic extra-galactic crossover energy.\\

\acknowledgments{
We would like to thank Birgit Eberle and Yvonne Wong for discussions,
and Zoltan Fodor and Sandor Katz for supporting our numerical analysis based on their
original computer codes.
This work has been partially supported by US NSF (Grant Nos. OPP-0236449,
PHY-0140407, PHY-0244507), US DoE (Grant Nos. DE-FG02-95ER40896, DE-FG05-85ER40226),
NASA-ATP 02-000-0151, and the Wisconsin Alumni Research Foundation.}


\begin{thebibliography}{99}
\bibitem{paramcosm}
We have checked that for a change of $z_{\rm min}$ to zero and
for $E_{i,{\rm max}}$ larger than the default $10^{12.5}$~GeV, the resulting cosmogenic
neutrino fluxes do not change within the energy range shown in Fig.~\ref{gxg_flux}.
For smaller $E_{i,{\rm max}}$, the predicted fluxes get even smaller.
\end{thebibliography}
\end{document}